\documentclass{aa}

\usepackage{txfonts}
\usepackage{graphicx}
\usepackage{natbib}

\usepackage{soul}

\newcommand{\ks}{$K_{S}$}
\newcommand{\arcdeg}{$^{\circ}$}
\newcommand{\xtesixteen}{XTE\,J1650$-$500}

\newcommand{\fwhm}{{\it fwhm}}
\newcommand{\lmxb}{LMXB}
\newcommand{\lmxbs}{LMXBs}
\newcommand{\psf}{{\it PSF}}

\begin{document}


\title{Disentangling the  NIR/optical  emission of\\ the black hole
  \xtesixteen\ during outburst\thanks{Based on observations made with
    the European Southern Observatory telescopes obtained from the
    ESO/ST-ECF Science Archive Facility.}
}


\titlerunning{}

\author{
  P.A.~Curran\inst{1}
  \and S.~Chaty\inst{1,2}
  \and J.A.~Zurita Heras\inst{3}
}


\institute{
  Laboratoire AIM (UMR 7158 CEA/DSM-CNRS-Universit\'e  Paris Diderot), Irfu/Service d'Astrophysique, CEA-Saclay, 91191, Gif-sur-Yvette Cedex, France 
  \and Institut Universitaire de France, 103, bd Saint-Michel, 75005 Paris, France
  \and Fran\c{c}ois Arago Centre, APC, Universit\'e Paris Diderot, CNRS/IN2P3, CEA/DSM, Observatoire de Paris, 13 rue Watt, 75205 Paris Cedex 13, France 
}

\date{Received ; accepted}


\abstract
{ While the sources of X-ray and radio emission in the different
  states of low-mass X-ray binaries are relatively well understood,
  the origin of the near-infrared (NIR) and optical emission is more
  often debated. It is likely that the NIR/optical flux
  originates from an amalgam of different emission regions, because 
  it occurs at the intersecting wavelengths of multiple processes. }
%
{ We aim to identify the NIR/optical emission region(s) of one such
  low-mass X-ray binary and black hole candidate, \xtesixteen, via
  photometric, timing, and spectral analyses.}
%
{ We present unique NIR/optical images and spectra, obtained
  with the ESO--New Technology Telescope, during the peak of the 2001
  outburst of \xtesixteen.  }
%
{ The data suggest that the NIR/optical flux is due to a combination
  of emission mechanisms including a significant contribution from
  X-ray reprocessing and, at early times in the hard state, a
  relativistic jet that is NIR/radio dim compared to similar
  sources. }
%
{ The jet of \xtesixteen\ is relatively weak compared to that of other black
  hole low-mass X-ray binaries, possibly because we observe as it
  is being ``turned off'' or quenched at the state transition.  While
  there are several outliers to the radio--X-ray correlation of the hard
  state of low-mass X-ray binaries, \xtesixteen\ is the first
  example of an outlier to the NIR/optical--X-ray correlation.}

\keywords{ 
  accretion, accretion disks 
  --  infrared: stars 
  -- ISM: jets and outflows 
  --  X-rays: binaries 
  -- X-rays: individuals: \xtesixteen
}

\maketitle

\section{Introduction}\label{section:intro}

Transient low-mass X-ray binaries (\lmxbs) are, for most of
the time, in a state of quiescence with faint or non-detected X-ray
emission and near-infrared (NIR)/optical emission dominated by the
main-sequence companion star (possibly with significant contribution
from the cold accretion disk).  They are often only discovered when ---
powered by an increased level of accretion onto the central compact
object (black hole or neutron star) --- there is a dramatic increase of
the X-ray, NIR/optical and radio flux.  During these outbursts the
systems are observed to go through several high-energy (X-ray)
spectral states before returning to a quiescent state, usually on
timescales of weeks, months, or even longer.


\lmxbs\ with a black hole compact object are observed (in X-rays)
initially in a generally low-intensity, power-law dominated, {\it
  hard} state before transitioning to a usually higher intensity,
{\it thermal-dominant}, {\it soft} state that decreases in flux and
evolves via a late hard states back into a quiescent state. Some
outbursts also display a {\it steep power law} state with a steeper
spectral slope than the regular hard state during the transition
between hard and soft states.
In general, the X-ray hardness versus intensity diagrams (HIDs) are
observed to follow canonical trajectories (e.g.,
\citealt{homan2001:ApJS132,homan_2005ApSS...300,belloni_2010LNP...794})
and are often (though not always) indicative of the black hole nature
of the compact object. Additionally, the hard states are associated
with aperiodic variability of the light curve not present in the soft
state and, in hard and steep power law states, quasi-periodic
oscillations (QPOs) are detected (for a fuller description of the
various possible states see \citealt{mclintock2006:csxs157}).

The origin of the NIR, optical and ultraviolet (UV) emission from a
black hole \lmxb\ is much less  understood than the origins of the
X-ray or radio emission because the optical wavelengths are at the
intersection of a number of different emission mechanisms (for reviews
of optical properties of \lmxbs\ see e.g.,
\citealt{vanParadijs1995:xrbi.nasa.58,Charles2006:csxs.book}).  The
companion star is generally relatively dim and may not contribute
significant flux during outburst when other mechanisms are
active. Intrinsic, thermal emission from the hot, outer accretion disk
may contribute at UV and optical wavelengths (e.g.,
\citealt{Shakura1973:A&A.24,Frank2002:apa.book}) though the
reprocessing of X-rays in the same region of the accretion disk is
thought to be a significant source of flux at wavelengths from UV
through to NIR (e.g.,
\citealt{Cunningham1976:ApJ.208,vanParadijs1994:A&A.290}). Recently,
evidence has been mounting that the relativistic jet, usually detected
in radio, also produces a significant contribution to the NIR flux, at
least in the hard state (e.g.,
\citealt{Jain2001:ApJ554,Corbel2002:ApJ.573,Chaty2003:MNRAS.346,Russell2006:MNRAS.371}),
and it is possible that the power law X-ray emission in the hard state
extends to, and contributes at, optical wavelengths. This emission is
ascribed to the inner regions of the system --- either a corona,
advection-dominated accretion flow, or the base of a compact jet ---
which we will refer to as the corona for simplicity.


The black hole X-ray transient \xtesixteen\ was initially detected by
the {\it Rossi X-ray Timing Explorer} (RXTE) on September 5 2001
\citep{Remillard2001:IAUC.7707}. In the following days, its variable
nature was confirmed
\citep{Remillard2001:IAUC.7707,Markwardt2001:IAUC.7707,Wijnands2001:IAUC.7715}
and quasi-periodic oscillations (QPOs) indicative of black hole
\lmxbs\ were observed \citep{Revnivtsev2001:IAUC.7715}. Optical
\citep{Castro-Tirado2001:IAUC.7707,
  Augusteijn2001:IAUC.7710,Buxton2001:IAUC.7715} and ATCA radio
\citep{Groot2001:IAUC.7708} counterparts were also reported at the
time. 
The canonical black hole state behaviour of the source over the $\sim
100$ day long outburst is well documented in both X-ray
\citep{Rossi2004:NuPhS.132} and radio \citep{Corbel2004:ApJ.617}, while
evidence of the black hole nature of the compact object is offered by
the confirmation of QPOs \citep{Homan2003:ApJ.586}, the detection of a
relativistic Fe K$\alpha$ line (e.g.,
\citealt{Miller2002:ATel.81,Miniutti2004:MNRAS.351}) and a derived
mass of $ \simeq 5.1 {\rm M}_{\odot}$ \citep{Slany2008:A&A492}.

The optical observations of the source during outburst (cited above)
were limited to a few epochs, but on June 10 2002, in quiescence,
\cite{Sanchez2002:IAUC.7989} obtained 15 optical spectra with FORS2 on
the 8.2m VLT at ESO, Paranal. Using these spectra and relative optical
photometry (obtained with the 6.5m Clay telescope at Las Campanas
Observatory, Atacama during May, June, and August 2002),
\cite{Orosz2004:ApJ616} derive an orbital period of 0.32 days  (7.69
hours). On August 2 2002 \cite{Garcia2002:ATel.104} estimated the
quiescent optical magnitude.
In this paper we present the only significant NIR/optical observations
during outburst, obtained by the ESO NTT in September and October 2001
(Table\,\ref{table:nights}), which comprises all available
unpublished, archived ESO data of the source.  In section
\ref{section:observations} we introduce the observations and reduction
methods, while in section \ref{section:results} we present the results
of our photometric, timing, and spectral analyses of the data. We
discuss the interpretation of our findings within the context of black
hole \lmxbs\ in section \ref{section:discussion} and summarise in
section \ref{section:conclusions}.


\section{Observations and reduction}\label{section:observations}

\subsection{Photometry}

\begin{figure} 
  \centering 
  \resizebox{\hsize}{!}{\includegraphics[angle=-0]{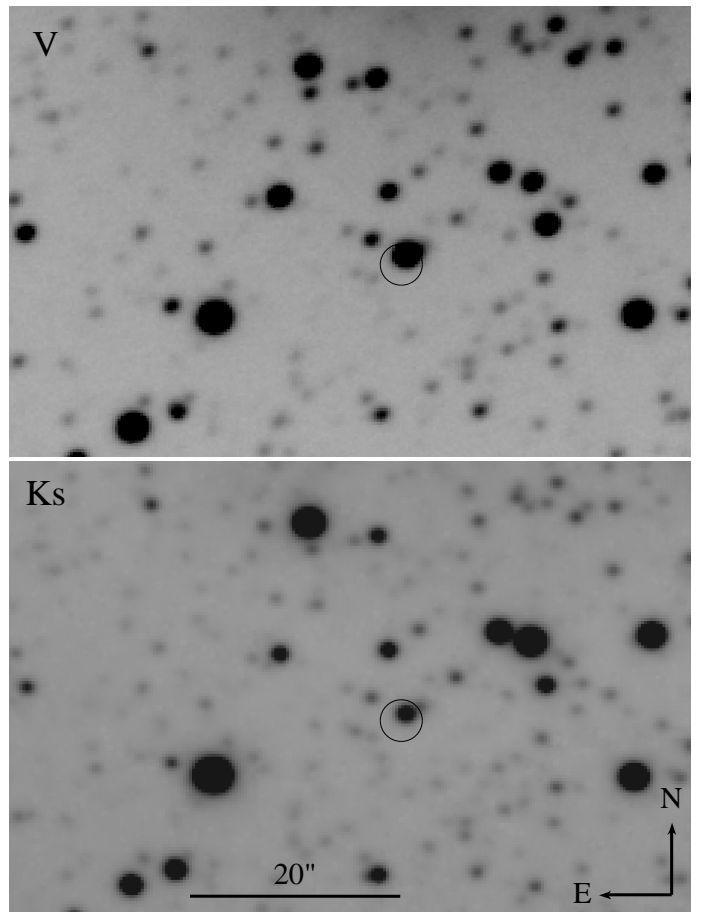}}
  \caption{NTT 60\arcsec\ $\times$ 40\arcsec\ finding charts
    ({\it upper:} 970s $V$ image; {\it lower:} 910s \ks\ image) with the
    2\arcsec\ optical positional uncertainty 
    \citep{Castro-Tirado2001:IAUC.7707} marked by a circle. }
  \label{fig.finding_chart} 
\end{figure}

\begin{table}	
  \centering	
  \caption{Nights of observations} 	
  \label{table:nights} 	
  \begin{tabular}{l l l l l} 
    \hline\hline
    Night  & MJD & Filters  & Size$^{\dagger}$($\arcmin$) & ESO ID\\ 
    \hline 
    2001-09-08   & 52161 & $J,H,K_{S}$ & $5\times5$ & 67.B-0486\\
    & & $GBF,GRF$* & ---  & 67.B-0486\\
    2001-09-14   & 52167 & $V,R,I$      & $9\times9$  & 67.B-0315\\ 
    2001-09-15   & 52168 & $RILD\#1$* & --- & 67.B-0486\\ 
    2001-09-24   & 52177 & $J,H,K_{S}$ & $4\times4$  & 67.D-0200\\
    2001-10-04          & 52187 & $J,H,K_{S}$ & $5\times5$  & 68.D-0316\\
    2001-10-06          & 52189 & $K_{S}$       & $4\times4$ &  68.D-0316\\
    2001-10-09          & 52192 & $V,R,I$      & $9\times9$ & 68.A-0440\\
    2001-10-10       & 52193 & $RILD\#1$* & ---  & 68.D-0144\\
    2001-10-13       & 52196 & $V$           & $1\times1$ & 68.D-0144\\ 
    2003-05-13            & 52772 & $I$            & $3\times10$ & 71.D-0337\\
    \hline 
  \end{tabular}
  \begin{list}{}{}
  \item[] $^{\dagger}$  Final image size after cropping; * Spectra.  
  \end{list}
\end{table}

Optical ($V,R,I$) and NIR ($J,H,K_{S}$) data were obtained 
with the ESO Multi-Mode Instrument (EMMI;
\citealt{Dekker1986:SPIE627}) and the Son of ISAAC (SofI) infrared
spectrograph and imaging camera on the $3.58$m ESO -- New Technology
Telescope (NTT), respectively, on a number of nights in September 
and October 2001, and in
May 2003 (Table\,\ref{table:nights}).
Data were reduced using the {\small IRAF} package, in which crosstalk
correction, bias-subtraction, flatfielding, sky subtraction, bad pixel
correction and frame addition were carried out as necessary.  
NIR data obtained on the nights of September 8 and October 4  did
not use a dithered pointing pattern as is usual for NIR images, therefore
it was not possible to produce a reliable sky for subtraction;
instead the sky produced on September 24 was used. This should be of
minimum effect because we used relative photometry to estimate magnitudes.

The images were astrometrically calibrated against 2MASS
\citep{Skrutskie2006:AJ.131} or USNO-B1.0 \citep{Monet2003:AJ125}
within the GAIA package.  The position of the source was derived via
the point spread function (\psf) photometry of the deep \ks\ band
image on October 6 (MJD 52188; seeing $\approx 1.2$\arcsec) as
16:50:00.95 $-$49:57:44.34 with a positional error\footnote{All
  uncertainties in this paper are given with a confidence of
  $1\sigma$.} dominated by the 0.1\arcsec\ 2MASS systematic
uncertainty (Figure\,\ref{fig.finding_chart}).

\begin{figure} 
  \centering 
  \resizebox{\hsize}{!}{\hspace{-15mm}\includegraphics[angle=-90]{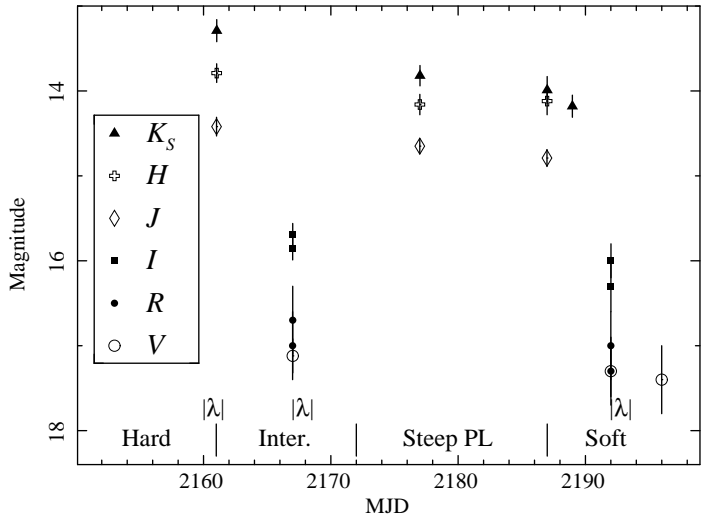}}
  \caption{$V,R,I,J,H$, and \ks\ band light curves (as presented in
    Table \ref{table:magnitudes}) with the periods of the hard,
    intermediate, steep power law and soft states
    \citep{Corbel2004:ApJ.617} marked. $|\lambda|$ marks  the epochs
    of the spectral observations.}
  \label{fig.lightcurve} 
\end{figure}

Relative \psf\ photometry was carried out on
the final images using the {\small DAOPHOT} package
\citep{stetson1987:PASP99} within {\small IRAF}.  However, the optical
data obtained on September 14 and October 9 were so poorly binned that
 producing an acceptable \psf\ model was impossible 
({\it full width at half maximum, \fwhm} 
$\lesssim2$ pixels); hence these images use relative
aperture photometry.
The NIR magnitudes (Table\,\ref{table:magnitudes};
Figure\,\ref{fig.lightcurve}) were calibrated
against the 2MASS catalogue  using $\approx 200$ objects per image,
after outliers were removed. The calibration, at least for the \ks\
band, was confirmed on the October 6 \ks\ image and four
\cite{Persson1998:AJ116} photometric standards, observed on the same
night (the only night on which reliable NIR standards were available).

Optical magnitudes (Table\,\ref{table:magnitudes};
Figure\,\ref{fig.lightcurve}) were calculated relative to $\approx$10
relatively isolated field stars, which were calibrated on the
September 14 images against \cite{Landolt1992:AJ.104} photometric
standards, observed on the same night (the only night on which optical
standards were obtained).  The errors on the optical magnitudes are
dominated by this absolute photometric calibration but unfortunately,
due to the saturation of USNO-B1.0 objects in the field, we are unable
to calibrate against that, or other, catalogues. The upper limit of the
late $I$ band image is approximated from the dimmest observable
object in the region of interest.
Because we were only able to obtain aperture photometry on
September 14 and October 9, the measured magnitude is likely to suffer
contamination from the unresolved, nearby sources (see
\cite{Orosz2004:ApJ616} for a better resolved image from the $6.5$m
Clay telescope).  To correct for this, we estimated the magnitude of
the contaminating sources to be $I = 19.2 \pm 0.5$ from the magnitude
at the position in the late $I$ band image (May 13 2003) when the
source had faded; $V = 18.4 \pm 0.4$ from the residual of the \psf\
photometry on October 13 2001; $R = 18.7 \pm0.4$ from an interpolation
of the $I$ and $V$ contaminating magnitudes. The difference in
magnitude ranges from $\gtrsim$1 magnitude for the $V$ band to
$\gtrsim$3 magnitudes for the $I$ band. In each case, given the
relatively large absolute calibration error, the corrected magnitude
(Table\,\ref{table:magnitudes}) is consistent, within errors, with the
magnitude uncorrected for contamination.

The data obtained on October 6 and October 13 consisted of multiple
high-cadence, ``fast'' photometry images, in \ks\ and $V$ bands,
respectively, to investigate possible short-term variability
(Table\,\ref{table:magnitudes}) up to $\sim$ 2--3 hours,
with an approximate sampling of an
image every 90s. \psf\ photometry was carried out on each of these
individual images, again using {\small DAOPHOT}. Source magnitudes
were calculated relative to a number of field stars (10 in NIR and 8
in optical) and normalized so that the weighted average is equal to
zero.  In each, a few magnitudes were eliminated due to
unsatisfactory \psf\ subtraction residuals which became obvious during visual
inspection of the images.

\begin{table}	
  \centering	
  \caption{Optical and NIR exposures and magnitudes} 	
  \label{table:magnitudes} 	
  \begin{tabular}{l l l l} 
    \hline\hline
    MJD-50000 & Filter & N $\times$ Exp (s) & Magnitude  \\ 
    \hline    
    2167.00354 & $V$  & $1\times30$  & 17.1 $\pm$ 0.2  \\
    2191.99546 & $V$  & $1\times120$ & 17.3 $\pm$ 0.3 \\
    2195.99182 & $V$  & $97\times10$ & 17.4  $\pm$ 0.4  \\
    2166.99704 & $R$  & $1\times150$ &  17.0 $\pm$ 0.4  \\
    2167.00545 & $R$  & $1\times30$   & 16.7 $\pm$ 0.4  \\
    2191.99848 & $R$  & $1\times120$ & 17.0 $\pm$ 0.4 \\
    2192.00531 & $R$  & $1\times40$   & 17.3 $\pm$ 0.4\\
    2167.00002 & $I$   & $1\times60$ & 15.86 $\pm$ 0.13 \\
    2167.00699 & $I$   & $1\times30$ & 15.69 $\pm$ 0.13   \\
    2192.00104 & $I$   & $1\times90$  & 16.3 $\pm$ 0.3 \\
    2192.00696 & $I$   & $1\times40$  & 16.0 $\pm$ 0.2 \\
    2772.17920 & $I$   & $4\times300$ & $>$19.0  \\
    2161.02124 & $J$  & $5\times20$  &  14.42 $\pm$ 0.11  \\
    2176.98651 & $J$  & $9\times10$   &  14.65 $\pm$ 0.09 \\
    2186.98230 & $J$  & $5\times20$ & 14.79 $\pm$ 0.10 \\
    2161.02907 & $H$ & $5\times20$  &  13.79 $\pm$ 0.11 \\
    2176.99397 & $H$  & $9\times10$  & 14.16 $\pm$ 0.12 \\
    2186.99013 & $H$  & $5\times20$ & 14.12 $\pm$ 0.16 \\
    2161.03691 & $K_{S}$ & $5\times20$  &   13.29 $\pm$0.13 \\
    2177.00145 & $K_{S}$ & $9\times10$  & 13.82 $\pm$ 0.12 \\
    2186.99797 & $K_{S}$ & $5\times20$ &  13.99 $\pm$ 0.16 \\
    2188.97400 & $K_{S}$ & $130\times7$ &  14.18 $\pm$ 0.13 \\
  \hline 
    2160.98738  & $GBF$ & $2\times180$  & --- \\
    2160.99906 & $GRF$ & $2\times360$  & ---\\
    2167.99490 & $RILD\#1$ & $2\times600$ &  ---\\
    2193.01287 & $RILD\#1$ & $2\times900$   & --- \\
\hline
\end{tabular}
\end{table}

\subsection{Spectral energy distributions}

 \begin{figure} 
  \centering 
  \resizebox{\hsize}{!}{\hspace{-15mm}\includegraphics[angle=-0]{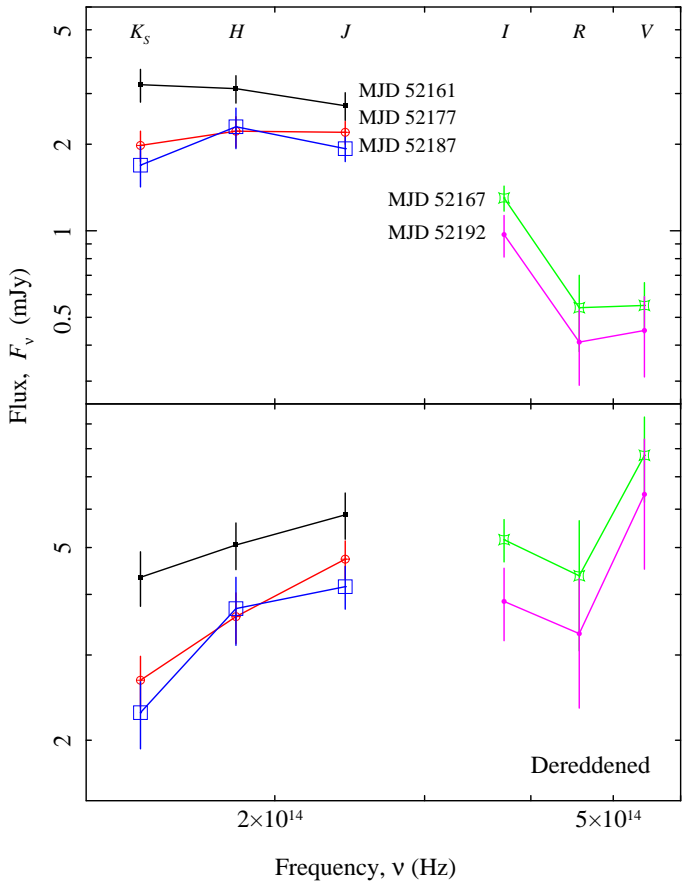}}
  \caption{Flux versus frequency plot at five different epochs,
    uncorrected (upper) and corrected (lower) for the Galactic
    extinction in the direction of the source of $E_{B-V} = 0.923$.}
  \label{fig.SED} 
\end{figure}

The observed magnitudes (Table\,\ref{table:magnitudes}) or, if
there were multiple observations per band per night, the
weighted average magnitudes, were first converted into flux densities,
$F_{\nu}$, at frequency $\nu$.  They were then converted into flux per
filter, $F_{filter}$ in units of photons\,cm$^{-2}$\,s$^{-1}$.  This
conversion was made via $F_{filter} = 1509.18896 F_{\nu}$ $( \Delta\lambda/\lambda
)$ where $\lambda$ and $\Delta\lambda$ are the effective wavelength
and  width of the filter in question.  {\tt XSPEC}
compatible files for spectral energy distribution (SED) fitting were
then produced from the flux per filter value using the {\small FTOOL},
{\tt flx2xsp}.
%
In figure \ref{fig.SED} we correct for the Galactic extinction in the
direction of the source \citep{schlegel1998:ApJ500} of $E_{B-V} =
0.923$ ($A_K \sim 0.3$, $A_V \sim 2.9$;
\citealt{cardelli1989:ApJ345})\footnote{This extinction
  should be treated with caution as estimates so close to the Galactic
  plane ($<5\deg$) are uncertain.}. This value of extinction is
consistent with that implied \citep{Guver2009:MNRAS.400} from the
X-ray absorption of this source \citep{Montanari2009:ApJ.692}, 
even though the source is likely closer \citep{Homan2006:MNRAS.366} and 
should not suffer the full effect of Galactic dust.

\subsection{Spectroscopy}

Optical and NIR spectra were obtained with EMMI and SofI on nights in
September and October 2001 (Tables\,\ref{table:nights},
\ref{table:magnitudes}). EMMI obtained red (3,850--10,000\,\AA)
low-dispersion spectra using grism\,\#1 ($RILD\,1$) on two
nights. SofI obtained usable blue ($GBF$; 9,500--16,400\,\AA) and red
($GRF$; 15,300--25,200\,\AA) low-resolution spectra on a single night,
while spectra from another two nights (September 18 and October 4
2001) were not useful due to a very high level of spatial dispersion,
as well as a lack of spectra for wavelength calibration on those
nights.
The data were reduced using the {\small IRAF} package in which
crosstalk correction, flatfielding and dark- or bias subtraction were
carried out as necessary.  To correct for NIR sky, the dithered NIR
exposures were subtracted from each other and the resulting spectra
summed later.

Spectra were extracted and reduced within the {\small IRAF} package,
{\tt noao.twodspec}.  Wavelength calibrations were performed against
helium + argon (optical) or xenon (NIR) lamps whose spectra were
extracted using the same parameters as for the relevant
source. Atmospheric telluric features significantly affect the
spectra and were corrected for by dividing the source spectrum by that
of a telluric standard at a similar airmass, using the {\tt telluric}
tool within {\small IRAF}. This procedure often causes artefacts in
the corrected spectra and, in the case of our optical spectra, these
artefacts  are significant, which is why we did not apply this procedure. 
Neither
the optical nor the NIR spectra were flux calibrated, but they were normalized.

\begin{figure*}
  \centering 
  \resizebox{\hsize}{!}{\includegraphics[angle=-90]{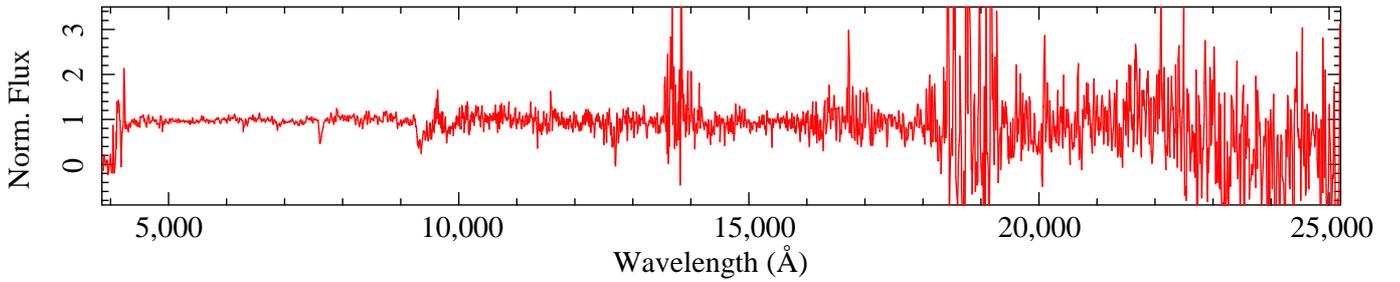}}
  \caption{Final normalized spectrum from 3,850 to 25,200\,\AA.  The
    apparent features just redward of 4,000\AA\ and blueward of
    $10,000$\,\AA\ are at the edge of the optical grism range where
    the response is not reliable. The high levels of noise at $\sim
    14,000$\,\AA\ and $\sim 19,000$\,\AA\ are
    artefacts of the telluric correction while the absorption feature
    at 7,625\,\AA\ is atmospheric.  }
  \label{fig.spectra} 
\end{figure*}

The NIR spectra (from the night of September 8 2001) were combined with
the optical spectrum from the night of September 15 2001 and the final
spectrum, from 3,850--25,200\,\AA, is plotted in Figure
\ref{fig.spectra}.  Owing to discrepancies in the optical wavelength
calibrations between the two nights, we are unable to sum the two
optical spectra and instead plot the spectra with the lower noise.  The
(wavelength-dependent) resolution of the final spectrum is 7--10\,\AA,
with a wavelength calibration error of $\lesssim 20$\,\AA\ (optical)
or $\lesssim 40$\,\AA\ (NIR).
The apparent features just redward of 4000\AA\ and blueward of
$10^{4}$\,\AA\ are at the edge of the optical grism range where the
response is not reliable. The high levels of noise at $\sim
14,000$\,\AA\ and $\sim 19,000$\,\AA\ are
artefacts of the correction for telluric features, while the general
trend towards higher noise at the red end is a real effect of the
observations.
An absorption feature with a \fwhm\ of 60\AA\ at 7,625\,\AA\ is
detected in the optical spectrum of the night of September 15, but this
is an atmospheric feature, also detected with similar properties in
the spectroscopic standard on that night.  This feature is not
detected in the October 10 spectrum, though this spectrum has a root
mean square noise about four times greater than the September 15
spectrum.

\section{Results}\label{section:results}

\subsection{High-cadence photometry}\label{section:fastphot}

\begin{figure} 
  \centering 
  \resizebox{\hsize}{!}{\includegraphics[angle=-90]{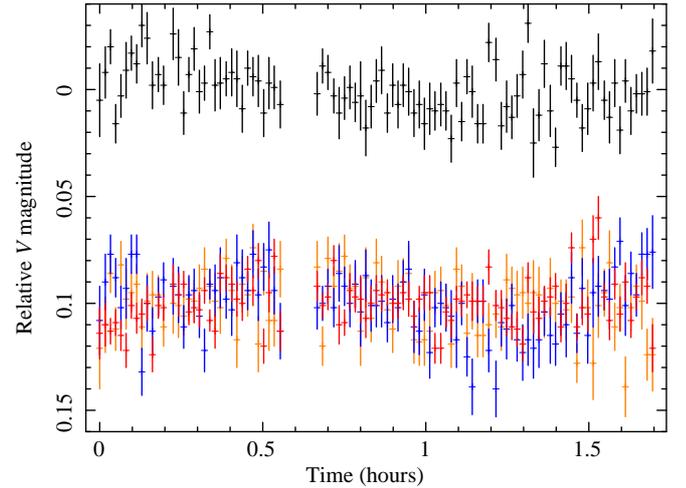}}
  \vspace{1mm}

  \resizebox{\hsize}{!}{\includegraphics[angle=-90]{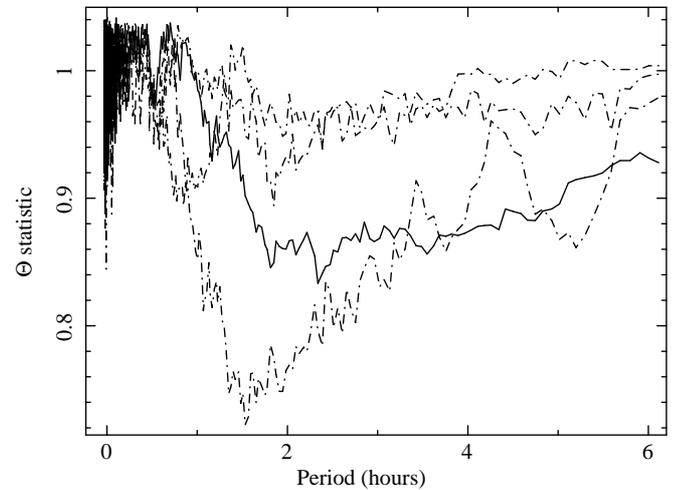}}
  \caption{{\it Top}: High-cadence $V$ band light curve of the source
    along with three comparison objects of a similar magnitude
    (normalized to a relative magnitude of 0.1) and, {\it bottom},
  the periodogram ($\Theta$ statistic versus period) for the source
  (solid line) and the nearby comparison objects (dashed lines).}
  \label{fig.V_lightcurves} 
\end{figure}

\begin{figure}
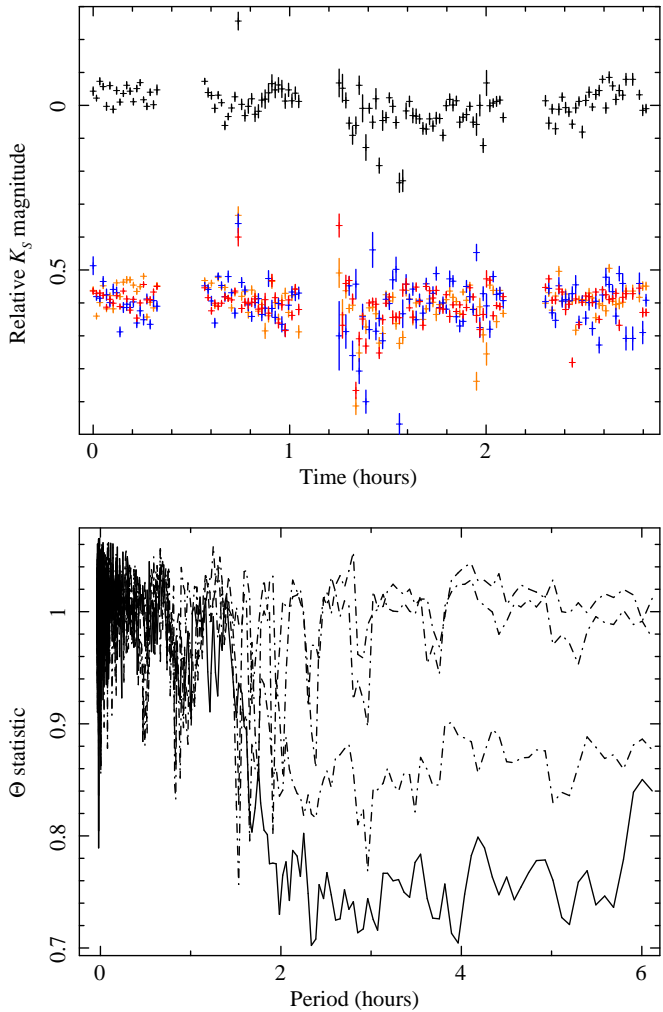
 
  \centering 
  \resizebox{\hsize}{!}{\includegraphics[angle=-90]{fig.K_photometry.ps}}
  \vspace{1mm}

  \resizebox{\hsize}{!}{\includegraphics[angle=-90]{fig.theta_K.ps}}
  \caption{{\it Top}: High-cadence \ks\ band light curve of the source
    along with three comparison objects of a similar magnitude
    (normalised to a relative magnitude of 0.6) and, {\it bottom},
  the periodogram ($\Theta$ statistic versus period) for the source
  (solid line) and the nearby comparison objects (dashed lines).}
  \label{fig.K_lightcurves} 
\end{figure}

Though the high-cadence light curves (Figures \ref{fig.V_lightcurves}
and \ref{fig.K_lightcurves}), particularly the \ks\ band, seem to
display some level of variability by eye, this is not supported by a
careful analysis of the data.
The standard deviation of the $V$ data is 0.012 magnitudes, compared
to an average error on the data points of $0.011 \pm 0.002$.  We fitted a
constant value to the light curve which, for the $V$ band, returns
$\chi^{2}_{\nu} = 143.1/(97-1) = 1.49$, statistically inconsistent
with being an acceptable fit at $>$99\% confidence. However, if
  we apply the same tests to a number of nearby sources of similar
  magnitude (and hence similar signal-to-noise ratio) we find similar
  standard deviations (0.011 -- 0.014) and $\chi^{2}_{\nu}$ (1.31 --
  2.05), for the same number of degrees of freedom.
This implies that any variability is not
intrinsic to the source itself but is likely noise.
Likewise, the \ks\ band data, which have a standard deviation of 0.059
magnitudes compared to an average error on the data points of $0.017
\pm 0.007$, returns $\chi^{2}_{\nu} = 1222/(128-1) = 9.6$,
inconsistent with a non-variable source.  However,several
  nearby sources of similar magnitude also return similar standard
  deviations (0.056 -- 0.086) and $\chi^{2}_{\nu}$ (9.1 -- 14.9),
  again implying that any variability is not intrinsic to the source
  itself and that, for the \ks\ band, the errors on the individual
  photometric points are underestimated, likely due to e.g. the
  difficulties of sky subtraction and \psf\ modelling in such a crowded
  field (Figure\,\ref{fig.finding_chart}).



Furthermore, we used the {\small IRAF} task, {\tt pdm} --- an
implementation of the phase dispersion minimization method of
\cite{Stellingwerf1978:ApJ224} --- to test for periodicity in the
variability of the light curves (see periodograms in lower panels of
figures \ref{fig.V_lightcurves} and \ref{fig.K_lightcurves}). We find
no reliable periodicity in either of the observed bands as the tests
returned Stellingwerf statistics, $\Theta \approx 1$ for all periods
less than the length of observations and broad minima of $\Theta
\gtrsim 0.7$ for longer test periods. While the $\Theta$ statistic in
each band was generally 
lower than that for the nearby comparison objects
(introduced above, dashed lines in periodograms), there was no strong
local minimum that would sugest a periodic variability. Neither is
there any sign of periodicity when the $\Theta$ statistic of the
source is normalized relative to that of the comparison objects.  This
null result was also confirmed using the Lomb-Scargle method for unevenly
sampled data \citep{Press1989:ApJ338}, which shows no significant
peak in power over the period ranges in question.

We also tested both light curves to see if either were consistent with
a sinusoidal periodicity similar to that of \cite{Orosz2004:ApJ616},
who measured a quiescent period of 0.3205 days (7.69 hours, compared to
the $\sim$2--3 hour duration of our observations) and an $R$ band
peak-to-peak amplitude of 0.2 magnitudes. This fit to the \ks\ light
curve returns $\chi^{2}_{\nu} = 7.2$ (125 degrees of freedom) while 
similar fits to the comparison objects return $\chi^{2}_{\nu}$ of
  between 10.3 and 14.2.  A similar fit to the $V$ band light curve of
 the source  returns $\chi^{2}_{\nu}
= 1.5$ (93 degrees of freedom) while fits to the comparison
  objects return $\chi^{2}_{\nu}$ of between 1.5 and 3.4. While it
would seem that both bands could be consistent with this period,
it is necessary in both cases that the epochs of observations fall
serendipitously at the minima of the periodic function, which seems
unlikely.
Indeed, to observe such a period in outburst would be
physically unrealistic since the periodicity was detected through
orbital eclipsing; during outburst, where the optical emission is 5
magnitudes brighter than quiescence ($V\sim24$, $R\sim22$;
\citealt{Garcia2002:ATel.104}), this would not be detected. However,
so-called ``superhump'' modulations, indicative of the orbital period,
have been observed in the hard state of several X-ray binaries
(e.g.,  \citealt{Zurita2008:ApJ.681}).

\subsection{Spectral energy distributions}\label{section:results:sed}

Though the light curves (Figure\,\ref{fig.lightcurve}) and flux-versus
-frequency plots (Figure\,\ref{fig.SED}) display apparent hints of NIR
colour evolution as well as a possible $V$ band excess, the spectral
indices ($\alpha$, where $F_{\nu} \propto \nu^{\alpha}$)
 of the two sets of $V,R,I$ data and the three sets of $J,H,K_{S}$
data are all consistent within the (not particularly well constrained)
errors.  Furthermore, a simultaneous fit of all five
(non-simultaneous) SED epochs does not support significant spectral
evolution, as the data are well fit  ($\chi^{2}_{\nu} = 0.72$ for 8
degrees of freedom) by an absorbed power law of spectral index,
$\alpha = 0.8 \pm 0.2$, with
the extinction set to $E_{B-V} = 0.923$, the Galactic value in that
direction \citep{schlegel1998:ApJ500}. The only evolution, at least
given the quality of the data, is that of the flux offset, which
decreases in time, as expected from the light curves. It is important
to note here that while spectral evolution is not statistically
supported, neither is it ruled out as a possibility.

Alternatively, the combined SEDs could also be acceptably fit by
simple black body radiation at any temperature $\gtrsim 3100$\,K
($\gtrsim 2.7\times10^{-4}$\,keV) with extinctions, $E_{B-V} \lesssim
1.6$, or at 7000\,K ($6\times10^{-4}$\,keV) for the Galactic
extinction ($\chi^{2}_{\nu} = 1.67$). Additionally, the SEDs are also
consistent with black body radiation from an accretion disk of
temperature $\gtrsim 5500$\,K ($\gtrsim 4.7\times10^{-4}$\,keV) with
extinctions, $E_{B-V} \lesssim 0.8$, or at $\approx 6\times 10^{6}$\,K
(0.5 \,keV) for the Galactic extinction ($\chi^{2}_{\nu} = 1.86$). It
is hence clear that the SEDs are not particularly constraining and
results should be treated with caution, especially because of the
time-averaged nature of the above fits.

\subsection{Spectroscopy}

The final spectrum (Figure \ref{fig.spectra}) displays a single
possible feature that we might attribute to the source: a tenuous
line (with a peak at $\sim 6$ times the noise and a \fwhm\ of 10\,\AA)
is detected at 16,700\,\AA.  The same feature is visible in the
spectrum before telluric correction, therefore is not an artefact of that
process. This line could correspond to an HI line at 16,810\AA\
(Brackett series, $\eta$), though given the difference in wavelength
and the very narrow \fwhm\ compared to the spectral resolution
(7--10\,\AA), it is likely that this line is not real. If this line
were real we would also expect to detect other lower energy Brackett
series lines that would have higher abundances, but we do not detect
any, again suggesting that the line may be a statistical fluctuation.
\cite{Augusteijn2001:IAUC.7710}, in their optical spectrum obtained on
September 8 2001 (on the night of our NIR spectra and seven nights prior
to our low-noise optical spectrum) identified Balmer series emission
lines, but we observe no such lines with a signal-to-noise ratio $\gtrsim 4$.
Neither do we detect the broad, double-peaked H$\alpha$ emission
  line from the accretion disk or the weak absorption lines from the
  companion star  identified by \cite{Sanchez2002:IAUC.7989} in their
VLT spectra from June 10 2002 (i.e., in quiescence).


\section{Discussion}\label{section:discussion}

\begin{figure*} 
  \centering 
  \resizebox{14cm}{!}{\includegraphics[angle=-90]{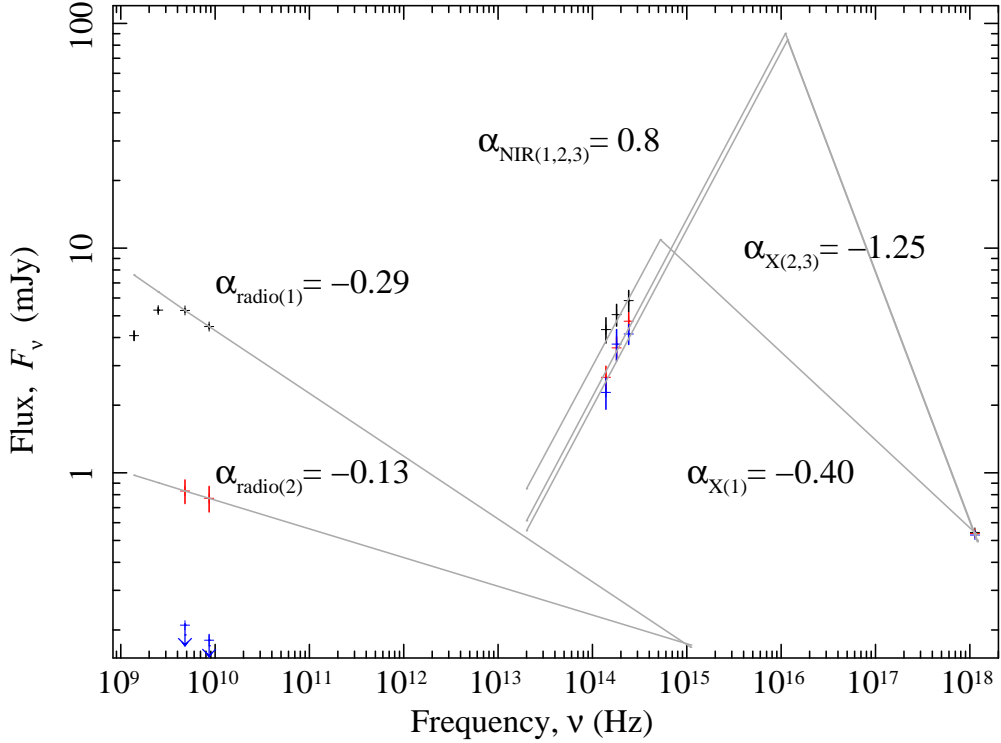}}
  \caption{Dereddened/unabsorbed (except for radio) flux-versus-frequency plot at three
    epochs (1 = MJD 52161, black; 2 = MJD 52177, red; 3 = MJD 52187,
    blue). Spectral indices, $\alpha$, are from previously published
    sources, except for the NIR. Power laws are purely
    phenomenological in order to show the relative behavior at
    different frequencies. }
  \label{fig.SED_all} 
\end{figure*}


The states of \xtesixteen\ are defined, from X-ray and radio
observations, as being hard up to MJD $\sim$52,161; hard/intermediate
up to MJD $\sim$52,172; steep power law up to MJD $\sim$52,187;
thermal-dominant/soft up to MJD $\sim$52,232; and hard after that date
\citep{Corbel2004:ApJ.617}.  Our observations were hence obtained in
the very late, initial hard/intermediate, steep power law and in the
first few days of the soft state
(Figure\,\ref{fig.lightcurve}). Specifically, our multi-band NIR
observations were obtained within 5 hours of observations 2, 3, and 4
of \citeauthor{Corbel2004:ApJ.617}, in the hard and steep power law
states. In figure\,\ref{fig.SED_all} the dereddened NIR and radio data
are plotted along with the X-ray flux from that paper (converted to
Jansky at 4.7\,keV). The common spectral index of the NIR observations
(section\,\ref{section:results:sed}) is also plotted along with the
fitted spectral indices of the radio (\citealt{Corbel2004:ApJ.617}
who used a thermal free-free absorbed power law) and
the X-ray \citep{Rossi2004:NuPhS.132} at those epochs.
One can see that epoch 1, in the hard state, is characterised by a
high level of radio emission and a relatively shallow-power-law
dominated X-ray flux, while epochs 2 and 3 display declining radio
flux and steep power-law-dominated X-ray emission.  One of the most
obvious aspects of this plot is that while the radio flux drops by
over an order of magnitude, the X-ray flux remains relatively stable,
dropping only fractionally. As pointed out by
\citeauthor{Corbel2004:ApJ.617}, this is consistent with the radio
originating from a self-absorbed compact jet while also indicating that
the jet contributes little, if any, flux to the X-ray
emission. Furthermore, the decrease to background of radio flux is
consistent with the assumed  behaviour of the jet in a black hole
system that is quenched during the soft, thermal state
\citep{Corbel2004:ApJ.617,Fender2004:MNRAS.355}.

\subsection{The source of NIR/optical emission}

It is expected that the NIR/optical emission of a black hole \lmxb\
originates from a combination of the companion star, the relativistic
jet, intrinsic thermal or reprocessed X-rays from the accretion disk,
and a possible contribution from the corona
(see section \ref{section:intro}).
The estimated quiescent optical magnitudes of this system ($V\sim24$,
$R\sim22$ \citealt{Garcia2002:ATel.104}; $I>19.0$ herein) are
significantly ($\sim$100--500 times) dimmer than those observed in
outburst and hence imply that the companion star has no measurable
effect on the emission during outburst.
If the observed NIR flux were due exclusively to the jet, 
we would expect an almost flat spectral index ($\alpha \approx 0$),
much shallower than that observed ($\alpha = 0.8 \pm 0.2$). 
We would also 
expect an approximately linear relationship between the NIR and radio
flux (which is assumed to be, and is consistent with being, due only
to jet emission). However, it is unclear how this relationship
  would evolve over a state transition, as is being observed here,
  since flux at one frequency may decrease earlier or faster than that
  at another frequency.  We find that the NIR flux falls off much
more gradually with $F_{{\rm NIR}} \propto F_{{\rm Radio}}^{0.21 \pm
  0.07}$ (at both radio frequencies), and maintains a high level of
emission even as the jet/radio emission goes to zero in the soft
state. Additionally, over the same time, the X-ray flux falls off even
less steeply than the NIR ($F_{{\rm NIR}} \propto F_{{\rm
    X}}^{25\pm10}$), therefore it seems plausible that at least some of the
NIR emission originates from the jet while the remaining emission must
be due to the accretion disk, reprocessing and/or coronal emission.

Unfortunately the X-ray flux cannot be extrapolated to imply an
accretion disk contribution to the NIR, as the accretion disk is not generally 
observed to make a significant contribution to the X-ray in the hard
or steep power law states. We would expect to observe spectral
emission lines if the NIR emission originated from an accretion disk,
but we observe none in our spectra during the hard/intermediate
states; however, this is not to say that the accretion disk is not
making a, possibly significant, contribution at these times.  Neither
can we extrapolate the X-ray flux to imply the corona's contribution
because, while it is clear that the power law contribution in the hard and
steep power law states must break at low energies, we cannot constrain
the form of this break.  Since the significant spectral change of the
X-ray does not seem to affect the NIR spectra, which remain constant,
one might be tempted to imply that the corona does not contribute to
the NIR, but because of the poorly constrained nature of the SED and the
lack of knowledge of the low-energy spectral shape of the corona, this
is not necessarily the case.
As mentioned earlier (section \ref{section:intro}), black hole \lmxbs\
display significant X-ray variability in the hard state at frequencies
from $\sim 10^{-3}$ to $10^{4}$\,Hz, which is absent from the soft
state, and this hard state variability has also been observed in the
optical from $\sim10^{-4}$ to $10$\,Hz (e.g.,
\citealt{Casella2010:MNRAS.404,Chaty2011:A&A.529}). The high level
of variability in both optical and X-ray is often attributed to the
non-thermal emission of the jet or corona.  We find no such
variability (section \ref{section:fastphot}) in our high-cadence optical and NIR observations (spanning
$\sim10^{-4}$ to $10^{-1}$\,Hz), which might be indicative of emission
associated with these regions, consistent with the source's soft state
at those epochs.

If the NIR emission was primarily due to X-ray reprocessing, the
observed NIR flux should be proportional to the observed X-ray flux as
$\propto F_{{\rm X}}^{\sim0.5}$ (\citealt{vanParadijs1994:A&A.290}, or
for more detailed calculations, \citealt{Coriat2009:MNRAS.400}) but
instead it follows the much steeper relationship of $F_{{\rm NIR}}
\propto F_{{\rm X}}^{25\pm10}$.  However, it must be noted here that
the level of reprocessing (and the NIR -- X-ray relationship) is very
sensitive to the X-ray spectral shape, and hence the X-ray range
sampled by a particular instrument. Over a state transition, as is
being observed here, the X-ray spectrum is undergoing dramatic changes
of spectral shape; consequently, while the X-ray flux over the measured range
remains constant, the flux over the range contributing to the
reprocessing may be significantly variable.
We do  find that the absolute
optical magnitude of the source  ($M_{V} = 2.2$) assuming a distance of
$2.6 \pm 0.7$\,kpc \citep{Homan2006:MNRAS.366},  agrees well with the
observed correlation with $\Sigma = (L_{{\rm X}} / L_{{\rm Edd}}
)^{1/2} P^{2/3} = 0.4$
\citep{vanParadijs1994:A&A.290,Deutsch2000:ApJ.530}, when we use the
observed period, $P = 7.69$\,days \citep{Orosz2004:ApJ616}, and the
derived mass of $ \simeq 5.1 {\rm M}_{\odot}$
\citep{Slany2008:A&A492}. This correlation was derived for
reprocessing, implying that the optical flux may originate from this
process. Furthermore, the lack of observed variability in the long
term optical --- in contrast to the NIR --- light curves
(Figure\,\ref{fig.lightcurve}) is consistent with the $F_{{\rm
    Optical}} \propto F_{{\rm X}}^{0.5}$ relationship expected from
reprocessing. The possible, though not statistically significant, $V$
band excess (Figure\,\ref{fig.SED}) may be due to reprocessing peaking
in that, or a bluer, band.

\subsection{The weak jet of \xtesixteen}

It has been noted \citep{Corbel2004:ApJ.617,Fender2010:MNRAS.406} that
\xtesixteen\ is X-ray loud, relative to the radio, when compared to
other black hole candidates (e.g.,
\citealt{Corbel2000:A&A.359,Corbel2003:A&A.400,Gallo2003:MNRAS.344,Fender2010:MNRAS.406,Coriat2011:MNRAS.414}). The
source also appears to be X-ray loud relative to the NIR, as in both
the hard (epoch 1) and the steep power law (epochs 2,3) states an
extrapolation of the X-ray spectra overestimates --- significantly in
the steep power law state --- the NIR flux, requiring a low-energy
break of the X-ray spectrum.  An excess X-ray emission is also
suggested by a comparison of the NIR and X-ray luminosities: assuming
a distance of $2.6 \pm 0.7$\,kpc \citep{Homan2006:MNRAS.366}, we find
NIR and X-ray luminosities in the hard state (epoch 1) of $L_{{\rm
    NIR}} \approx 10^{33}$\,erg\,s$^{-1}$ and $L_{{\rm X}} \approx
10^{37}$\,erg\,s$^{-1}$ (each with a 40\% error, dominated by the
uncertainty on distance). These luminosities fall under the \lmxb\
(black hole) hard state correlation of \cite{Russell2006:MNRAS.371},
implying that either the source is X-ray loud (by $\gtrsim 3$ orders
of magnitude) or that the source is NIR dim (by $\gtrsim 2$ orders of
magnitude). For these luminosities to fall on the correlation, the
distance would have to be a factor of 30, or more, greater ($\sim 80$\,kpc),
which is of course unphysical compared to the size of the Galaxy ($\sim
35$\,kpc diameter); indeed, considering the source's position in the
Galactic plane ($l,b = 336.7$\arcdeg\ $-3.4$\arcdeg) any distance much
greater than 8.5\,kpc is unlikely.
In the soft state (which has luminosities similar to those of the hard
state) the source appears to be NIR and optically dim relative to the distribution
of other soft state \lmxbs\ (\citeauthor{Russell2006:MNRAS.371}), but
not significantly so. Furthermore, as noted by
\cite{Gallo2008:ApJ.683}, the late-time quiescent X-ray luminosity
($L_{{\rm X}} \approx 3 \times 10^{30}$\,erg\,s$^{-1}$) is at the
lower end of the distribution of quiescent luminosities for black hole
binaries, though, when compared with the quiescent optical luminosity
($L_{{\rm Optical}} \approx 10^{30}$\,erg\,s$^{-1}$), the X-ray
luminosity falls on the correlation of
\citeauthor{Russell2006:MNRAS.371}, albeit at the low end. Obviously,
because of NIR/optical observational bias (i.e., we only observe the
brightest sources), caution has to be exercised when implying that a
source is of a low luminosity compared to a distribution as we do here
for the soft and quiescent states. However, this observational bias
should not affect our comparison to the hard state correlation.
Given that the hard state X-ray luminosity of this source is very
similar to the range of luminosities presented in
\citeauthor{Russell2006:MNRAS.371}
($10^{35}$--$10^{38}$\,erg\,s$^{-1}$), while the NIR luminosity is at
the low end of the observed range (at the quiescent level of other
\lmxbs), it seems more likely that the source is NIR, and hence radio,
dim (at least quantitatively) as opposed to X-ray loud, as suggested by
\cite{Corbel2004:ApJ.617}.


While both the NIR and radio fall below their respective hard state
correlations with the X-ray (in fact, this is the first example of an
outlier to the NIR/optical correlation), the calculated NIR
luminosities are more consistent with the correlation for the hard
states of neutron star \lmxbs\ \citep{Russell2006:MNRAS.371} --- in
contrast to the canonical black hole behaviour of the observed X-ray
properties of this source (section\,\ref{section:intro}). Likewise,
the radio luminosities of the source fall on the {\it efficient}
branch of the radio--X-ray correlation and are similar to the
luminosities of neutron stars \citep{Coriat2011:MNRAS.414}. The
accretion flow in the hard state of black hole binaries is generally
radiatively inefficient and as such a significant amount of the
gravitational potential may be converted to relativistic jet flow
(i.e., kinematically efficient; e.g., \citealt{Gallo2005:Natur.436}),
although that energy may also pass unobserved across the black hole's
event horizon, advected onto the compact object
\citep{Narayan1996:ApJ.457}.  Conversely, the accretion flow in
neutron star binaries (and the soft state of black holes) is
radiatively efficient and thus thermal accretion emission is a more
effective method of energy radiation than a jet (i.e., kinematically
inefficient, see e.g., \citealt{Fender2003:MNRAS.343}). Hence, it is
likely that the detected jet of \xtesixteen\ is weak due to
radiatively efficient accretion (this is what
\citeauthor{Coriat2011:MNRAS.414} refer to, qualitatively, as X-ray
loud). This is also consistent with the radio
\citep{Corbel2004:ApJ.617} and X-ray
\citep{Rossi2004:NuPhS.132,Montanari2009:ApJ.692,Corbel2004:ApJ.617}
light curves, which peak (except at very soft energies, where the soft
state thermal component is dominant) by the time of our first
observation, and it is consistent with the X-ray definitions of state since
our first observation is at the very end of the initial hard state.
Thus what we observe in both radio and NIR is a weak jet that
  may be in the process of being ``turned off'' or quenched due to
the state transition from hard to intermediate as the accretion flow
transitions from geometrically thick, optically thin and radiatively
inefficient (but kinematically efficient) to geometrically thin,
optically thick and radiatively efficient.


\section{Conclusions}\label{section:conclusions}

The data  of \xtesixteen\ suggest that the jet 
contributes to the NIR flux, at least at early times, before being
quenched completely by the time of the later, steep power law state
observation. However, we find that the NIR flux falls off much more
gradually than the radio and maintains a high level of emission even
as the jet/radio emission goes to zero --- suggesting that there is
also significant contribution from X-ray reprocessing and possibly the
accretion disk and/or corona, though we cannot distinguish which of
these might dominate due to a poorly constrained SED and a lack of
indicative features of either, such as spectral lines or photometric
variability.

The observed radio and NIR luminosities during the hard state are
orders of magnitude lower than expected for a source of this X-ray
luminosity, and while several outliers to the radio--X-ray
correlation of the hard state of \lmxbs\ exist, this is the first such
example of an outlier to the NIR/optical--X-ray correlation.  The
NIR/radio dim emission suggests that the jet of \xtesixteen\ is much
weaker than in the hard states of other black hole \lmxbs, possibly
because  we observe the jet as it is being ``turned off'',
or quenched, at the state transition.

\begin{acknowledgements}  
  We thank S. Corbel for useful discussions on the manuscript
  and the anonymous referee for constructive comments. 
  This work was supported by the Centre National d'Etudes Spatiales
  (CNES) and is based on observations obtained with MINE: the
  Multi-wavelength INTEGRAL NEtwork.  This research has made use of
  NASA's Astrophysics Data System and the SIMBAD database, operated at
  CDS, Strasbourg, France.
\end{acknowledgements}

\end{document}